\documentclass[preprint,showpacs,preprintnumbers,amsmath,amssymb]{revtex4}
\usepackage{graphicx}
\usepackage{dcolumn}
\usepackage{bm}
\usepackage{epsfig}
\begin{document}

\preprint{PREPRINT}

\title{Thermodynamic and Dynamic Anomalies for a Three
Dimensional Isotropic Core-Softened Potential}

\author{ Alan Barros de Oliveira}
\affiliation{ Universidade Federal do Rio Grande do Sul, Caixa 
Postal 15051, 91501-970, Porto Alegre, RS, BRAZIL}

\author{Paulo A. Netz}
\affiliation{ Departamento de Qu\'{\i}mica, ULBRA, Canoas 
RS, BRAZIL and Departamento de Qu\'{\i}mica, Unilasalle, Canoas 
RS, BRAZIL}

\author{ Thiago Colla}
\affiliation{ Universidade Federal do Rio Grande do Sul, 
Caixa Postal 15051, 91501-970, Porto Alegre, RS, BRAZIL}

\author{ Marcia C. Barbosa}
\affiliation{ Universidade Federal do Rio Grande do Sul, Caixa 
Postal 15051, 91501-970, Porto Alegre, RS, BRAZIL} 
\email{marcia.barbosa@ufrgs.br}

\date{\today}

\begin{abstract}

Using molecular dynamics simulations and integral
equations (Rogers-Young, Percus-Yevick and hypernetted chain closures)
we investigate the thermodynamic
of particles interacting  with  continuous core-softened intermolecular 
potential. Dynamic properties are also analyzed by the simulations.
We show that, for a chosen shape of the potential, the density, 
at constant pressure, has a maximum for a  certain temperature. The line 
of temperatures of maximum density (TMD) was determined in the pressure-temperature
phase diagram. Similarly the diffusion constant at a constant temperature, $D$, 
has a maximum at a density $\rho_{max}$ and a minimum at a density 
$\rho_{min}<\rho_{max}$. In
the pressure-temperature phase-diagram
the line of extrema in diffusivity 
is outside of TMD line. 
Although in this interparticle potential lacks directionality,
this is the same behavior observed in SPC/E water.

\end{abstract}


\maketitle

\section{Introduction}

Water is an anomalous substance in many respects.
Most liquids contract upon cooling. This is not the case of water, a liquid
where the specific volume at ambient pressure starts to increase 
when cooled below $T=4 ^oC$ \cite{Wa64}. Besides, in a certain
range of pressures, also exhibits an anomalous increase of compressibility 
and specific heat upon cooling \cite{Pr87}-\cite{Ha84}. Far less 
known are its dynamics anomalies: while for most
materials diffusivity decreases with increasing pressure,
liquid water has an opposite behavior in a large region
of the phase diagram 
\cite{St99,Ga96,Ha97,Sc91,Er01,Ne01,Ne02a,Ne02b,Ne02}.
The increase of diffusivity of water as the pressure
is increased is related to the competition between the local
ordered tetrahedral structure of the first neighbors and the distortions
of the structure of the first and second neighbors. In the region
of the phase diagram where this ordered structure is dominant,
increasing pressure implies breaking first neighbors
hydrogen bonds what allow for interstitial second neighbors to be in a closer
approach. The interactions are thus weakened and therefore,
although the  system  is more dense it has a larger
mobility. In this sense, a good model for water and tetrahedral 
liquids should not only
exhibit thermodynamic but also dynamic anomalies.
In SPC/E water, the region of the pressure-temperature (p -T) 
phase-diagram where the density anomaly appears is contained 
within the region of the p -T phase-diagram where anomalies in 
the diffusivity are present  \cite{Er01,Ne01}.

For explaining the thermodynamic anomalies, it was proposed
that these anomalies are related to a second critical
point between two liquid phases, a low density liquid
(LDL) and a high density liquid (HDL) \cite{Po92} located at the 
supercooled region beyond  the line of
homogeneous nucleation and thus it cannot be experimentally measured.

Water, however, is not 
an isolated case. There are also other examples of  tetrahedrally
bonded molecular liquids such as phosphorus \cite{Ka00,Mo03}
and amorphous  silica \cite{La00} that also are good candidates for having
two liquid phases. Moreover, other materials such as liquid metals
\cite{Cu81} and graphite \cite{To97} also exhibit thermodynamic anomalies.
Unfortunately a coherent and general interpretation of 
the low density liquid and high density liquid phases is still missing.

What type of potential would be appropriated for 
describing the tetrahedrally bonded 
molecular liquids? Directional interactions are  certainly an 
important ingredient in obtaining a quantitative  predictions
for network-forming liquids like water. However, the models
that are obtained  from that approach are too complicated, being
impossible to go beyond mean field analysis.
Isotropic models became the simplest framework to understand the physics of 
the liquid-liquid phase transition and liquid state anomalies. 
From the desire of constructing a 
simple two-body isotropic potential capable of describing
the complicated  behavior present in water-like molecules, 
a number of models in which
single component systems of particles interact via 
core-softened (CS) potentials \cite{pabloreview} have been proposed. 
They possess a 
repulsive core that exhibits a region of 
softening where the slope changes dramatically. This region can 
be a shoulder or a ramp 
\cite{St98,Sc00,Fr01,Bu02,Bu03,Sk04,Fr02,Ba04,Ol05,He05,He70,Ja98,Wi02,Ma04,Ku04,Xu05}.

In the first  case, the potential consists of a hard
core, a square repulsive shoulder and, in some cases, an attractive square well
\cite{St98,Sc00,Fr01,Bu02,Bu03,Sk04,Fr02,Ba04,Ol05,He05,Ma04,Wi02}. In two
dimensions, such potentials have density and diffusion anomalies
and in some cases a second critical point \cite{Sc00,Ba04,Ol05,He05}. In 
three dimensions, these potentials do not have 
dynamic and thermodynamic anomalies but possess a second 
\cite{Sk04} and sometimes a third \cite{Bu03} critical
point, accessible by simulations in the region
predicted by the hypernetted chain integral 
equation \cite{Fr01,Fr02,Ma04}.
 
In the second case, the interaction potential has  two competing
equilibrium distances, defined by a repulsive ramp \cite{Ja98,Wi02,Ku04}. By 
including a global term for attractions, this model
displays a liquid phase with a first-order line of liquid-gas 
transition ending in a critical point
\cite{Ja98,Xu05} brings this second critical point in to an 
accessible region of higher temperature, and also displays a 
normal gas-liquid critical point.

Notwithstanding the progresses made by the models described above, a 
potential in which both the potential and the force are continuous
functions and that exhibits all the thermodynamic and dynamic anomalies present
in water is still missing. 
In this paper, we check if particles interacting with a 
core-softened potential similar to the one proposed by 
Cho \emph{et al.} \cite{Ch96,Ch97a,Ne04}
exhibit thermodynamic and dynamic anomalies similar to the
ones present in water. Since the potential can have 
a variety of shapes, depending on its parameters, we 
study a soft ramp ( with continuous force) with
two  scale distances. This type of potential
gives a distribution function similar
to the one expected for SPC/E water \cite{Go93}.  We check if 
the region in the pressure-temperature phase-diagram of thermodynamic
anomalies is inside the region of dynamic anomalies as in SPC/E water
\cite{Ne01}.  

The reminder of this paper goes as follows.
In sec. II the model is introduced; in sec. III the phase-diagram is
obtained within the Rogers-Young, Percus-Yevick and hypernetted chain 
integral equations.
Results for the phase-diagram and for 
the diffusion constant obtained 
from molecular dynamics simulations are shown in sec. IV.
Conclusions about the relation between the locus of 
the density anomaly and the diffusion anomaly are presented in sec. V.

\section{The model}

We consider
a set of molecules of diameter $\sigma$ interacting
through a potential that consists of a combination of a 
Lennard-Jones
potential of well depth $\epsilon$ 
plus a Gaussian well centered on radius $r=r_{0}$ with depth $a$
and width $c$, 
\begin{equation}
U(r)=4\epsilon\left[\left(\frac{\sigma}{r}\right)^{12}-
\left(\frac{\sigma}{r}\right)^{6}\right]+
a\epsilon
\exp\left[-\frac{1}{c^{2}}\left(\frac{r-r_{0}}{\sigma}
\right)^{2}\right].
\label{eq:potential}
\end{equation}

This potential can represent a whole family of
two length scales  intermolecular
       interactions, from a deep double
       wells potential \cite{Ch96,Ne04} to a 
repulsive shoulder \cite{Ja98}, depending on the choice of the values
       of $a$, $r_{0}$ and $c$. Specific choices 
of these parameters leads to  double 
well potentials similar to the one studied by Cho et al.
 \cite{Ch96}.
The  attractive double well brings both the liquid-gas phase-transition
and the anomalies to  higher temperatures into the unstable region
of the p -T phase diagram \cite{Ne04}.

\begin{figure}[htb]
\begin{center}\includegraphics[clip=true,scale=0.5]{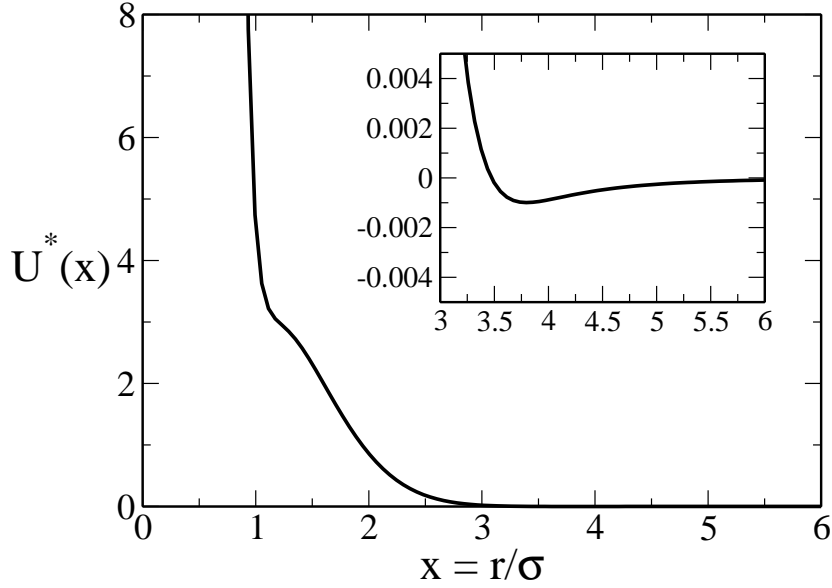}
\end{center}
\caption{Interaction potential eq. (\ref{eq:potential}) with
parameters $a=5,$ $r_{0}/\sigma=0.7$ and $c=1$, in  reduced
units. The inset shows a zoom in the very small attractive part 
of the potential
\label{cap:Soft-core}}
\end{figure}

In order to circumvent this 
difficulty, here  we investigate   the thermodynamic and dynamic behavior of
 particles interacting
via a potential with a very small attractive
region. We use   Eq.~ (\ref{eq:potential}) with
$a=5,$ $r_{0}/\sigma=0.7$ and $c=1$. 
This potential  has  two length scales within 
a repulsive ramp followed by a very small attractive well
( Fig. (\ref{cap:Soft-core}) ).

In order to have an overview of the behavior of particles interacting with this potential,
we use integral equations to estimate the thermodynamic properties in the phase diagram.

\section{Integral equations}

\begin{figure}[htb]
\includegraphics[clip,scale=0.6]{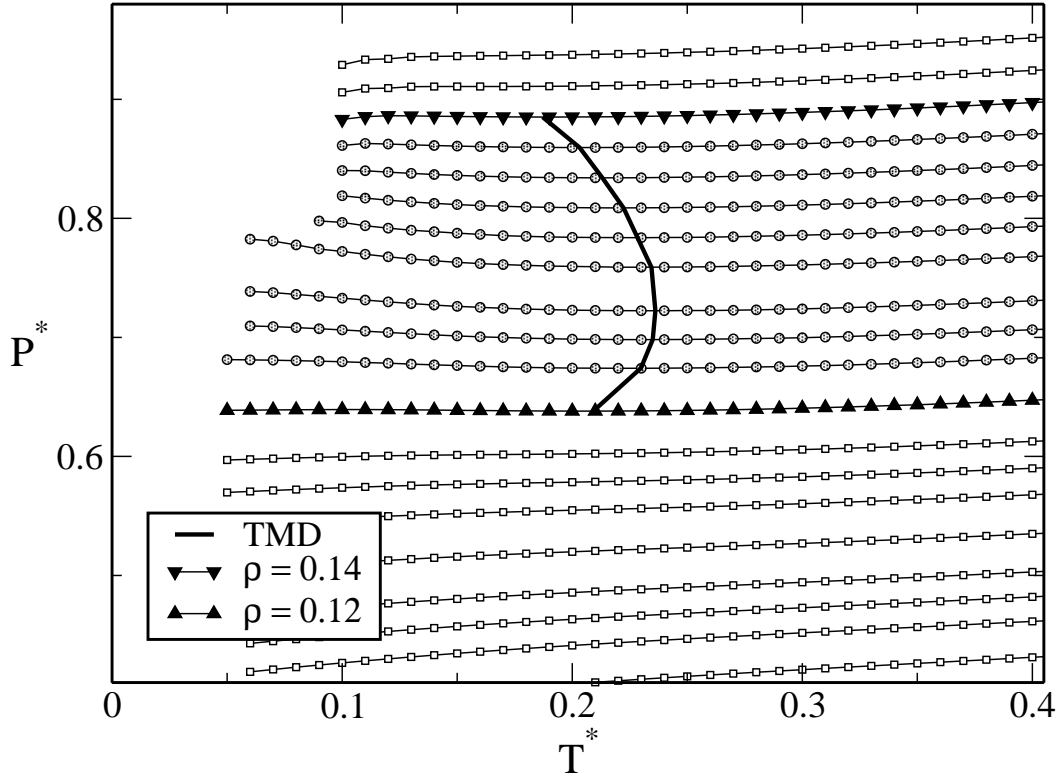}
\caption{Pressure-temperature phase-diagram obtained
by Rogers-Young integral equations. From
bottom to top, the isochores
$\rho = 0.100 ; 0.103 ; 0.105 ; 0.107 ;$
$0.110 ; 0.113 ; 0.115 ; 0.117 ; 0.120 ; $
$0.123 ; 0.125 ; 0.127 ; 0.130 ; 0.132 ; $
$ 0.134 ; 0.136 ; 0.138 ; 0.140 ; 0.142 $ and $0.144$  are shown.
The solid line illustrate the TMD. 
\label{cap:Phase-diagram}}
\end{figure}

\begin{figure}[htb]
\includegraphics[clip,scale=0.6]{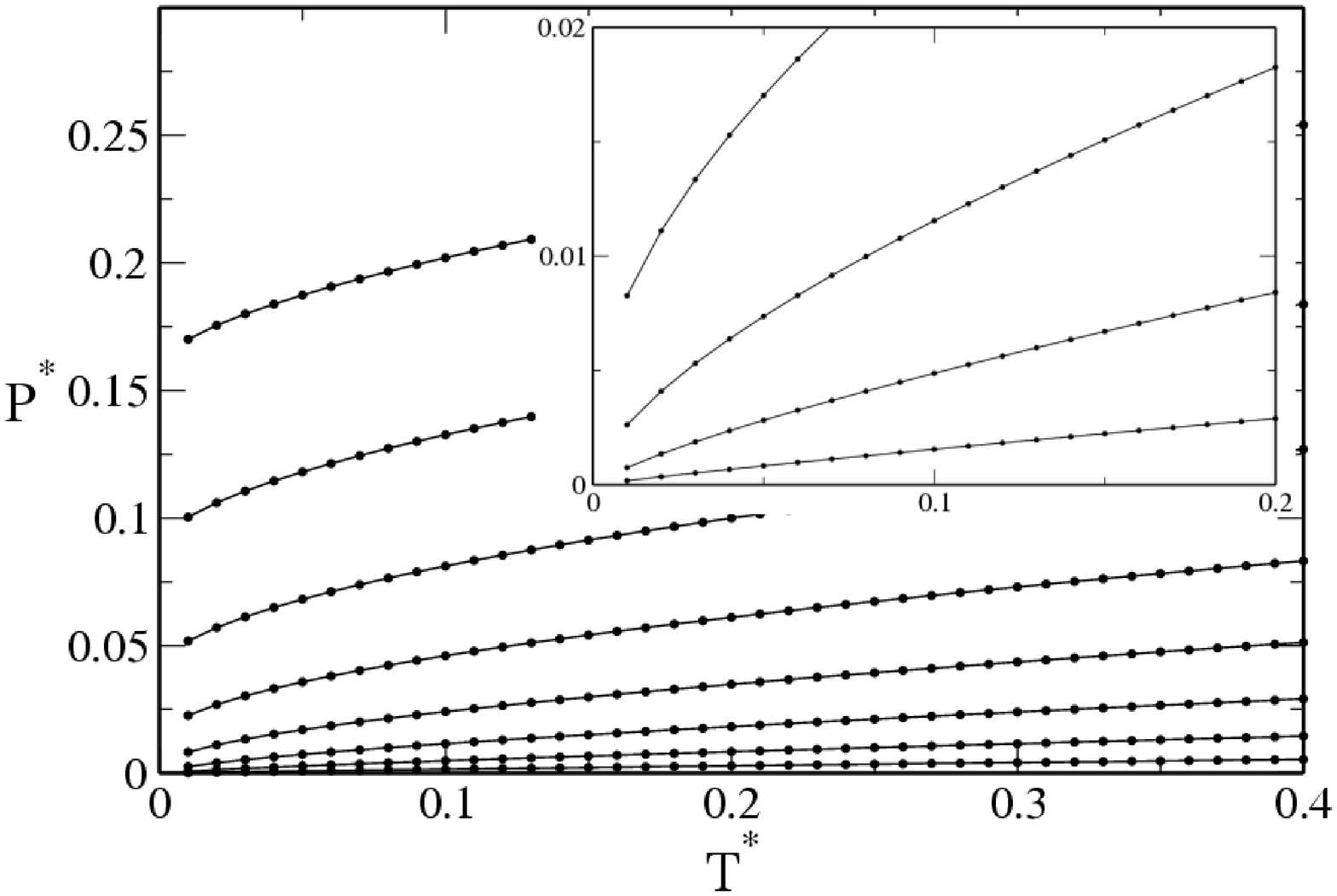}
\caption{Phase-diagram in Rogers-Young integral equations. 
The low isochores
$\rho = 0.01 ; 0.02 ; 0.03 ; 0.04 ;$
$0.05 ; 0.06 ; 0.07$ and $0.08$, from bottom
to top, are shown. The inset shows the isochores
for $\rho^{*}\leq0.03$ converging to a point
at supercooled region - the liquid-gas 
phase transition.
\label{cap:Zoom}}
\end{figure}

One of the most successful theories for describing the structure of
simple fluids are the integral equations \cite{Ha86}. Among
them, certainly the most famous is the Ornstein-Zernike (OZ)
equation \cite{Or14}
that, for pure isotropic fluids with density $\rho$, gives
an exact relation between the direct correlation function, c(r), and
the total correlation function, h(r), and it is given by

\begin{equation}
\gamma(r)=h(r)-c(r)=\rho\int h(\mathbf{r})c(|\mathbf{r-r^{'}}|)d\mathbf{r},
\label{eq:oz}
\end{equation}

where $h(r)=g(r)-1$ and where  $g(r)$
is the pair distribution function. $g(r)$ is proportional to the
probability to find a particle at a  distance $r$ when another particle is 
placed at the origin. 

The Fourier transform of Eq. (\ref{eq:oz}) is  given by
\begin{equation}
\Gamma(k)=\frac{\rho C(k)^{2}}{1-\rho C(k)},
\label{eq:ozft}
\end{equation}
where $\Gamma(k)$ and $C(k)$ are the Fourier
Transforms of $\gamma(r)$  and $c(r)$, and the 
definition for the direct correlation function 
\begin{equation}
c(r)=h(r)-\ln\left\{ g(r)\exp\left[\beta U(r)\right]\right\} 
+B(r),
\label{eq:dcf}
\end{equation}
was used. Here
$\beta=1/k_{B}T$ and $B(r)$ is the sum of
all bridge diagrams for the interparticle potential. Eq. (\ref{eq:ozft}) 
together with 
Eq. (\ref{eq:dcf})  can be solved
for a given interparticle potential. For obtaining 
$B(r)$  many approximations (\emph{closure}
relations)  have  been proposed \cite{Pe58,Le59,Le66,Ve80,Ma83,Ba86,Le92} 
along the years. Unfortunately, these approximations
have the following thermodynamic inconsistence: the pressure
calculated via the \emph{fluctuations route,}
\begin{equation}
\beta P_{fluc.}=\rho-4\pi\int_{0}^{\rho}\rho '\int_{0}^{\infty}r^{2}c(r,\rho ')drd\rho ',
\label{eq:pc}
\end{equation}
differs from the pressure  calculated via the \emph{virial route},
\begin{eqnarray}
\beta P_{vir.} = \rho -
\frac{2 \pi}{3}\rho^{2}\int_{0}^{\infty}r^{3}\frac{dU(r)}{dr}g(r)dr.
\label{eq:pv}
\end{eqnarray}

Two of these  closures have been widely 
used: the Percus-Yevick (PY)\cite{Pe58} where
\begin{eqnarray}
B(r)=\ln[1+\gamma(r)]-\gamma(r) \; ,
\label{eq:bpy}
\end{eqnarray}
and the hypernetted chain (HNC)\cite{Le59} that sets
\begin{eqnarray}
B(r)=0 \; .
\label{eq:bhnc}
\end{eqnarray}

While HNC is appropriated for 
large interparticle
distances,  PY is more adequate  for small ones. 
In order to avoid the inconsistencies present
in the original integral equations, Rogers and Young \cite{RY84} proposed 
a mix of the HNC and
PY closures of the form
\begin{equation}
c(r)=\exp[-\beta U(r)]
\left[1+\frac{\exp[\gamma(r)f(r)]-1}{f(r)}\right]-\gamma(r)-1,
\label{eq:ry}
\end{equation}
with the mixing function $f(r)=1-\exp[-\alpha r]$.
Note that at $r=0$  Eq. (\ref{eq:ry}) reduces to the PY
approximation and for  $r\rightarrow\infty$,
Eq. (\ref{eq:ry})  tends to the HNC approximation.
The Rogers-Young (RY) approximation puts together 
PY and HNC closures with an adjustable
parameter $\alpha$.  This parameter is determined by
imposing that the pressure calculated using Eq. (\ref{eq:pc}) gives
the same result as using Eq. (\ref{eq:pv})  (\emph{global} 
consistency criterion). 
This method have the inconvenience of the integral in $\rho '$.
Instead of calculating $\alpha$ by 
imposing that the pressure should be the same when
calculated using Eq. (\ref{eq:pc}) and Eq. (\ref{eq:pv}), one can 
obtain $\alpha$ by checking the consistency between the
compressibilities  $\chi_{fluc}$ and  $\chi_{vir}$,
calculated by derivation of Eqs. (\ref{eq:pc}) and 
(\ref{eq:pv}) respectively \cite{Ze86}, namely

\begin{equation}
\frac{\beta}{\rho}\chi_{fluc.}^{-1}=1-4\pi\rho\int_{0}^{\infty}r^{2}c(r)dr
\label{eq:chic}
\end{equation}
and
\begin{eqnarray}
\frac{\beta}{\rho}\chi_{vir.}^{-1} & = & 1-\frac{4\pi}{3}\rho\beta\int_{0}^{\infty}r^{3}
\frac{dU(r)}{dr}g(r)dr \nonumber \\
&  & -\frac{2\pi}{3}\rho²\beta\int_{0}^{\infty}r^{3}\frac{dU(r)}{dr}\frac{\partial g(\rho,r)}{\partial\rho}dr.
 \label{eq:chiv}\end{eqnarray}

Others closures  were  proposed  where one \cite{Ze86,Bo03} or more adjustable
parameters \cite{Ha80,Ma87,Le01} are needed in order to guarantee the 
consistency. In this work, we use the RY approximation
due  its  success
in describing the structure of the systems whose particles interact
by a purely repulsive pair potentials \cite{RY84,La99,Li98,Ku04}.

A numerical iterative solution of the system formed by Eq. (\ref{eq:ozft})
and Eq. (\ref{eq:ry}) was performed using a fine grid with $M=4096$
points and a step size $\Delta x=0.0075$, from $x=r/\sigma=0.0075$
until $x=M\Delta x$.  The tolerance for thermodynamic consistency was
$1-\chi_{fluc.}/\chi_{vir.}<10^{-3}$. For the PY and HNC closures, 
the same $M$ and $\Delta x$ was used, in the same
range. Pressure, temperature and density are shown in dimensionless
units:
\begin{equation}
\label{T*}
T^{*}\equiv \frac{k_{B}T}{\epsilon}
\end{equation}
\begin{equation}
\label{rho*}
\rho^{*}\equiv\rho \sigma^{3}
\end{equation}
\begin{equation}
\label{P*}
P^*\equiv\frac{P \sigma^{3}}{\epsilon}
\end{equation}

The main features of the phase diagram obtained by RY closure are illustrated
in Fig.~(\ref{cap:Phase-diagram}). This p -T phase-diagram
shows that the  isochores with $0.120\leq\rho^*\leq0.140$  have minimum 
which means that $(\partial P/\partial T)_{\rho}$ = 0. From 
this follows $(\partial \rho/\partial T)_{P}$ = 0,
which implies a density anomaly. The line of minima for the different
isochores forms the Temperatures of Maximum Density (TMD) shown in 
 Fig.~(\ref{cap:Phase-diagram}) by a solid line. 
 
The presence of a possible
critical point between two liquid phases 
may be suggested  by the crossing of the analytic 
continuation of isochores $\rho^{*}=0.134;0.136;0.138;0.140;0.142;0.144$, 
in the region below $T^{*}<0.05$. In this region the integral equations 
numerical solutions do not converge and the thermodynamic equilibrium 
is not achieved for the MD simulations.

Fig. (\ref{cap:Zoom}) shows that the very low isochores at
$\rho^{*}\leq0.03$
are converging to a point in the supercooled region, 
indicating a liquid-gas critical point, as can be seen
from the inset.

\begin{figure}[htb]
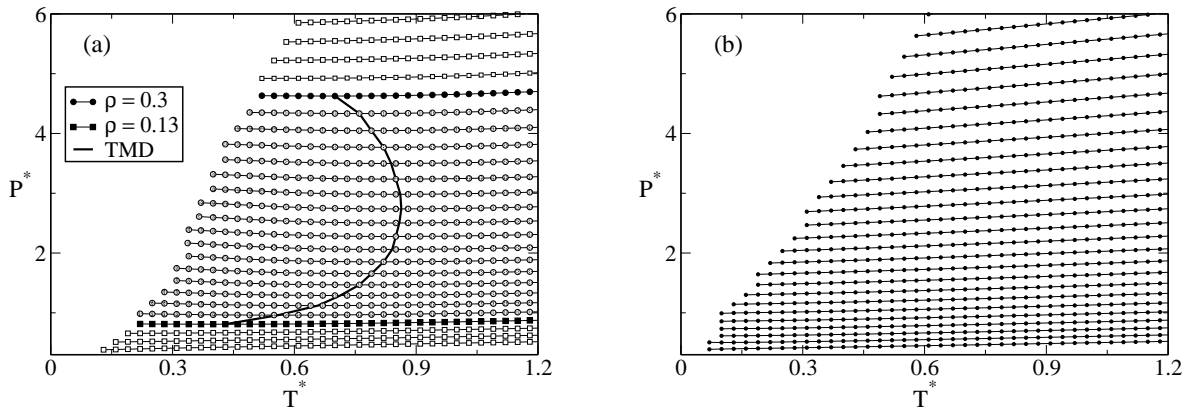

\center
\includegraphics[clip=true,scale=0.3]{phase-diagram-py.eps}
\hspace{1cm}\includegraphics[clip,scale=0.3]{phase-diagram-hnc.eps}
\caption{p -T phase-diagram obtained
by PY (Fig.(\ref{cap:pyhnc}a)) and HNC (Fig.(\ref{cap:pyhnc}b)) 
integral equations. From bottom to top, the twenty five isochores 
illustrated are $\rho = 0.10; 0.11; 0.12; ...; 0.33; 0.34.$ in both 
figures. The solid line in Fig. (a) illustrates the TMD line.
 \label{cap:pyhnc}}
 \end{figure}

When analyzing the model Eq.~(\ref{eq:potential}) with the PY approximation, 
density anomaly was found between $0.13<\rho^*<0.3$  in a region 
of temperatures from $0.4<T^*<0.86$, as can be seen from Fig. (\ref{cap:pyhnc}a). 
No density anomaly was found when 
the model Eq.~(\ref{eq:potential}) was analyzed with HNC (Fig. (\ref{cap:pyhnc}b)).

\section{The Molecular Dynamics}

\begin{figure}[htb]
\includegraphics[clip,scale=0.6]{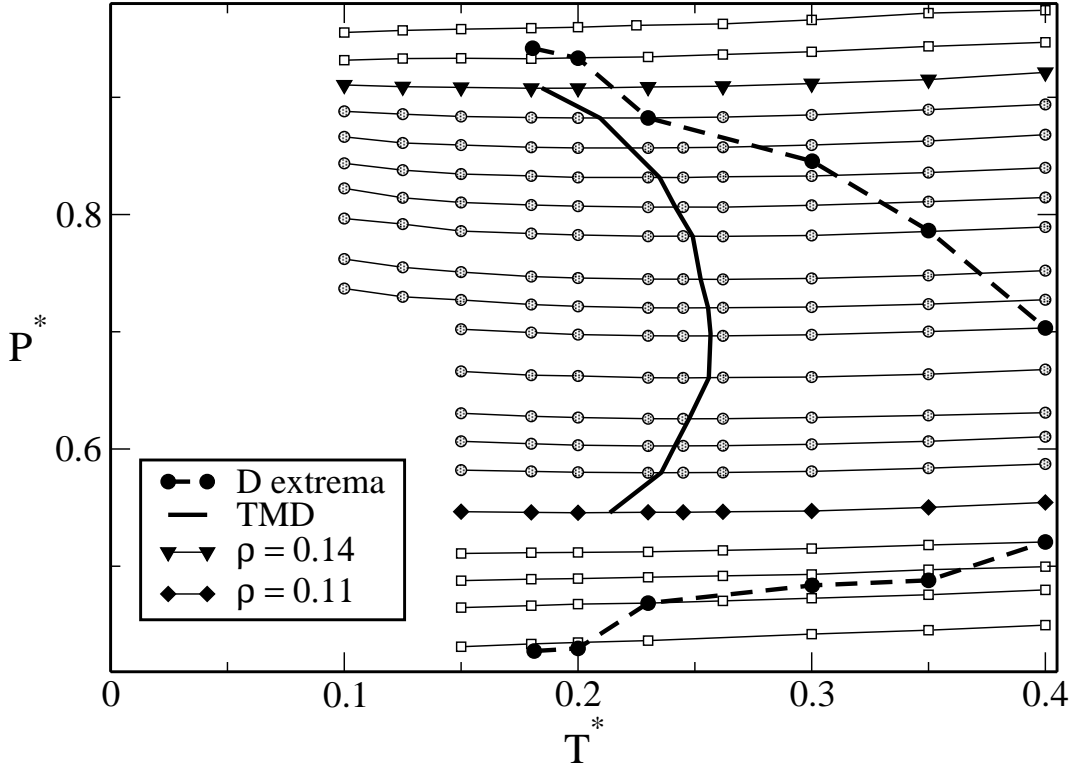}
\caption{Pressure-temperature phase-diagram obtained
by molecular dynamic simulation. From
bottom to top, the same isochores illustrated in RY
phase-diagram,
$\rho = 0.100 ; 0.103 ; 0.105 ; 0.107$
$0.110 ; 0.113 ; 0.115 ; 0.117 ; 0.120 ; 0.123 ; $
$0.125 ; 0.127 ; 0.130 ; 0.132 ; 0.134 ; 0.136 ; $
$0.138 ; 0.140 ; 0.142$ and $0.144$ 
are shown.
The solid line illustrates the TMD and the dashed 
line shows the boundary of the diffusivity extrema.
 \label{cap:md}}
\end{figure}

\begin{figure}[htb]
\includegraphics[clip,scale=0.6]{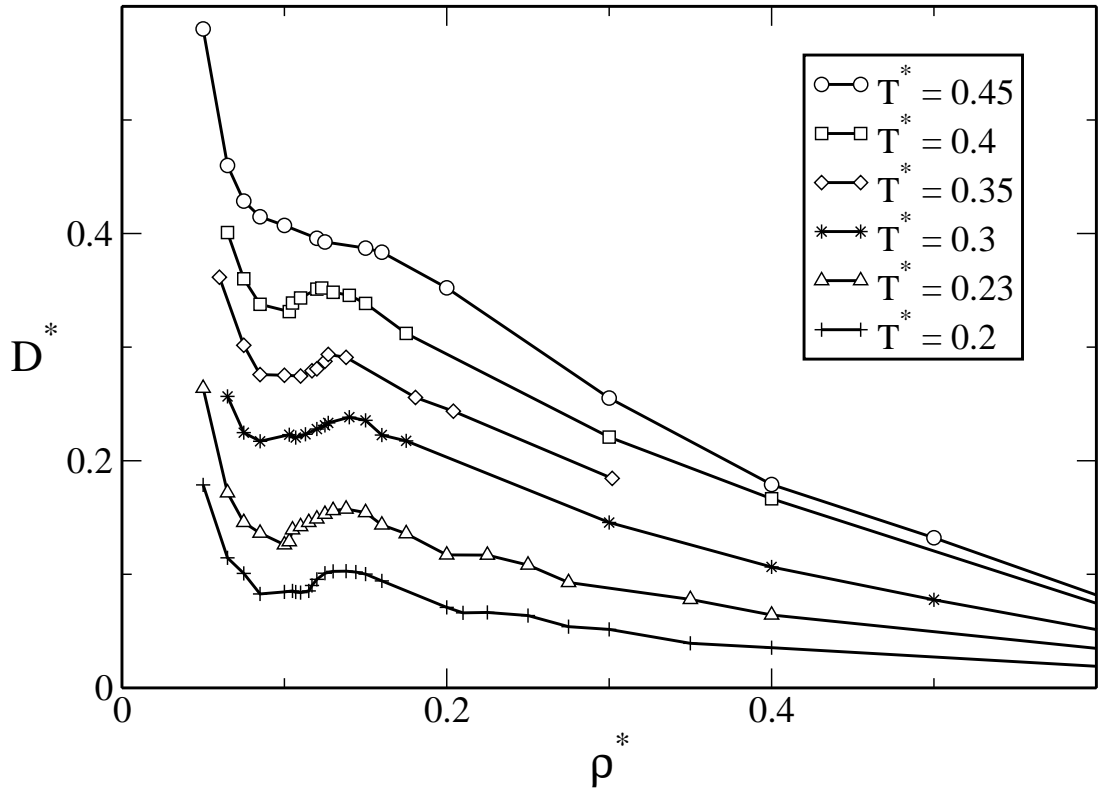}
\caption{ Diffusion coefficient as a function of
density for some studied temperatures. The units are
defined in the text.
\label{cap:D}}
\end{figure}

We also 
performed molecular dynamics simulations in the canonical ensemble
using 500 particles in a cubic box with periodic boundary conditions, 
interacting with the intermolecular potential described above. 
The chosen parameters were $a = 5.0$, $r_{0}/\sigma = 0.7$ and $c = 1.0$.
The cutoff radius was set to 3.5 length units. 
Using reduced units defined as $T^{*}$ and
$\rho^{*}$, a broad range of temperatures 
(0.10 $\leq$ $T^{*}$ $\leq$ 0.45) and 
densities (0.05 $\leq$ $\rho^{*}$ $\leq$ 1.00)
was chosen, in order to explore the phase diagram.
Thermodynamic and dynamic properties were calculated over 2 500 000 steps
long simulations, previously equilibrated over 500 000 steps. In the 
lower temperature systems, additional simulations were carried out
with equilibration over 2 000 000 steps, followed by  6 000 000 simulation
run. The time step was 0.001 in reduced units. 
The thermodynamical stability of the system was checked 
analyzing the dependence of pressure on density and also by
visual analysis of the final structure, searching for cavitation. 

Figure (\ref{cap:md}) shows the p -T phase-diagram obtained by
molecular dynamics. The isochores have minima that define the 
temperature of maximum density. The TMD line encloses the region of density
(and entropy) anomaly. The comparison between the RY and MD results shows
that the TMD line starts at lower densities in the MD simulations
than that in RY integral equations. Above $\rho^{*}=0.144$ 
both theories agree that no density anomaly happens. 
The RY pressures for each isochore
are slightly underestimated when compared with simulations,
but the overall agreement between the predictions of this
closure and the simulations results is very good.
On the other side, the PY approximation
predicts density anomaly, but in a region completely different than that of MD.
The HNC closure  do not shows TMD line, as we have discussed before.

The MD simulations also indicates the possibility of a
liquid-liquid critical point by the crossing of the analytic
continuation of isochores $\rho^{*}=0.134;0.136;0.138;0.140;0.142;0.144$ -
the same behavior was seen by the RY closure, and missing by PY and HNC.

We also study the mobility associated with the potential described in
Eq.~(\ref{eq:potential}).  The
diffusion is calculated using the the
mean-square displacement averaged over different initial times,
\begin{equation}
\langle \Delta r(t)^{2} \rangle = \langle [r(t_0+t)-r(t_0)]^2\rangle\; .
\end{equation}
Then the diffusion coefficient is obtained from the relation
\begin{equation}
D=\lim_{t\to\infty}\langle \Delta r(t)^{2} \rangle/6t \; .
\end{equation}

Figure (\ref{cap:D}) shows the behavior of the  translational diffusion
coefficient, 
\begin{equation}
\label{D*}
D^*\equiv\frac{D(m/\epsilon)^{1/2}}{\sigma}
\end{equation}
as a function of $\rho^*$.
At low temperatures, the behavior
is similar to the behavior found \cite{Ne01} in SPC/E supercooled
water. The diffusivity increases as the density is lowered, reaches a
maximum at $\rho_{D{\rm max}}$ (and $P_{D{\rm max}}$) and decreases
until it reaches a minimum at $\rho_{D{\rm min}}$ (and $P_{D{\rm min}}$).

The region in the p -T plane where there is an anomalous behavior in
 the diffusion is bounded by $(T_{D{\rm min}},P_{D{\rm min}})$ and
 $(T_{D{\rm max}},P_{D{\rm max}})$ and their location is shown
in Fig.~ (\ref{cap:md}).
The region of diffusion anomalies $(T_{D{\rm
 max}},P_{D{\rm max}})$ and $(T_{D{\rm min}},P_{D{\rm min}})$ 
lies outside the region of density anomalies like in SPC/E water \cite{Ne01}.

\section{Conclusions}

We have studied the thermodynamic properties of fluids interacting
via a three dimensional continuous core-softened potential with a
continuous force, using several integral equations closures, RY, PY and HNC, 
as well as molecular dynamics simulations. The continuity of the 
force is similar that one expect for realistic systems. We studied the density
anomaly and anomalies in  the translational diffusion.
Both  RY integral equations  and molecular
dynamics results show that the density can behave anomalously at a
certain range of pressures and temperatures. The agreement 
between these two theories is very good, confirming
the RY integral equations as a powerful 
tool for investigations of interatomic repulsive
pair potentials. The PY approximation emphasizes the short-ranged
interactions and indeed 
predicts density anomaly, but the width of the anomaly region
is strongly overestimated if compared 
with MD simulations. No density anomaly was found employing
the HNC closure, because this approach is better suited for systems
with long-ranged interactions.
  
  Both MD and RY theories suggests the possibility 
of a second critical point, between two liquid phases,
by the crossing of the analytic continuation of the isochores 
where $0.134\leq\rho^{*}\leq0.144$ for $T^{*}<0.05$. However
the actual calculations or simulations in this region was not
possible, either by failure in the integral equations convergence
and because the equilibrium was not reached.

The translational diffusion shows a maximum and a minimum 
in the pressure-temperature phase-diagram.
The region in the p -T plane of
density anomaly   is located inside
the region of the anomalous diffusion.

The studied continuous  core-softened potential,
despite of not having long-ranged or H-bond-like
directional interactions,
exhibit thermodynamic and dynamic anomalies 
similar to the ones observed in SPC/E water \cite{Ne01}.

\subsection*{Acknowledgments}

We thank to Anatol Malijevsky for having introduced us 
to the computational methods to the integral equations,
Giancarlo Franzese for fruitful discussions 
about the potential
and the Brazilian science agencies CNPq, FINEP  and Fapergs 
for financial support.

\end{document}